\documentclass[secnumarabic,amssym,amsmath,amsfonts,nobibnotes,twoside,aip,jmp]{revtex4}

\newcommand{\C}{\ensuremath{\mathbb{C}}}

\begin{document}

\title{\bf Complex Hadamard matrices for prime numbers }
\author{Petre Di\c t\u a}
\email{dita@zeus.theory.nipne.ro}

\affiliation{ Horia Hulubei National Institute of Physics and Nuclear Engineering, P.O. Box MG6, Magurele,  Romania }

\begin{abstract}
In this paper we disprove the Haagerup statement that all complex Hadamard matrices of order five are equivalent with the Fourier matrix $F_5$ by constructing
  circulant matrices that lead to  new Hadamard matrices.  An important item is the construction of  new mutually unbiased bases that are a basic concept of quantum theory and play an essential role in quantum tomography, quantum cryptografy, teleportation, construction of dense coding schemes, classical signal proccesing, etc.
\end{abstract}

\maketitle

\section{Introduction}

Bj\"orck G. and Fr\"oberg R.,  \cite{BF}, seem to be the first authors who treated the problem of cyclic n-roots with applications to Hadamard matrices.

 T. Durt and his coworkers, \cite{D}, studied the classification problem of Hadamard matrices of size $ n \le 5$. In particular they used the {\em dephased} form for all matrices. They says that the (rescaled) Fourier matrices are  the unique example in order $n \le 3 $, and in order 5 one has uniqueness again, result which already is absolutely non-trivial! Their example for $n\,=\,3$ is the following
 \begin{eqnarray} 
F_3= \left[\begin{array}{ccc}
1&1&1\\*[2mm]
1&\gamma&\gamma^2\\*[2mm]
1&\gamma^2&\gamma
\end{array}\right]\label{f3}\end{eqnarray}
where $\gamma\,=\,e^{\frac{2\pi i}{3}}$, citing the paper  \cite{BF}. 

 Because this approach is spreading fast, see paper \cite{BT}, we construct many three and five dimensional Hadamard matrices that disprove the Sz\"oll\H osi assumption that only the Fourier matrices $F_3$ and  $F_5$ have a real  existence. 
Sz\"oll\H osi in his thesis, \cite{S}, seems to agree that a complete classification of complex Hadamard matrices is only available up to order $n\,=\,5$, and in this case it is equivalent with the Fourier matrix $F_5$.
 
In this paper we consider the cases $n\,=\,2$, $n\,=\,3$ and $n\,=\,5$.  

The orthogonality concept is essential for getting new complex Hadamard matrices and in the following we make use of the particular class of inverse orthogonal matrices, $O = (o_{ij})$, whose inverse is given by  
\begin{eqnarray} 
O^{-1}=(1/o_{ij}^t)=(1/o_{ji})\label{orth}\end{eqnarray}
where $t$ means transpose, and their entries, $0\ne o_{ij} \in \C $,
satisfy the relation
\begin{eqnarray} O O^{-1} = n I_n\label{di}\end{eqnarray}
When $ o_{ij}$ entries take unimodular values, $O^{-1}$ coincides with the Hermitean conjugate $O^{*}$ of $O$, and in this case $O/\sqrt{n}$ is the definition of complex Hadamard matrices, see for example  paper \cite{JJS}.

\section{The two dimensional case}

We start with the simplest case $n\,=\,2$  and we use the next matrix in order to find new Hadamard matrices
 \begin{eqnarray} 
C_2= \left[\begin{array}{cc}
a&b\\*[2mm]
c&d
\end{array}\right]\label{c2}\end{eqnarray}

This matrix is not Hadamard. Thus we make use of the relation (\ref{di}) which provides the following constraint $ b c+ a d\,=\,0$.  By using it we find four solutions 
\begin{eqnarray}\begin{array}{cccc}
H_1=\left[\begin{array}{cc}
\frac{b\,c}{d}&b\\*[2mm]
c&d
\end{array}\right], &
H_2=\left[\begin{array}{cc}
a&-\frac{a\,d}{c}\\*[2mm]
c&d
\end{array}\right], &
H_3=\left[\begin{array}{cc}
a&b\\*[2mm]
-\frac{b\,c}{a}&d
\end{array}\right], &
H_4=\left[\begin{array}{cc}
a&b\\*[2mm]
c&-\frac{b\,c}{a}
\end{array}\right]\label{h14}\end{array}\end{eqnarray}

The above matrices generate many MUbs ({$\mathbb{I}$, $H_{1}$, $H_{2}$}), ({$\mathbb{I}$, $H_{1}$, $H_{3}$}),({$\mathbb{I}$, $H_{2}$, $H_{4}$}), etc.

\section{The three dimensional case}

As usual we make use of the Sylvester ortogonality, see paper \cite{JJS}, and we start with the circulant matrix $C_3$ whose form is 
 \begin{eqnarray}
C_3= \left(\begin{array}{ccc}
a&b&c\\*[2mm]
c&a&b\\*[2mm]
b&c&a
\end{array}\right)\label{c3}\end{eqnarray}
and $C_3$ matrix provides the following parameter constraints

\begin{eqnarray} a^2b+b^2c+ac^2\,=\,0, \,\,ab^2+a^2c+bc^2\,\,=\,\,0 \label{eqs}\end{eqnarray}

When they are satisfied  $C_3 $ transforms  into new matrices which are not yet Hadamard. For example from the first equation 
(\ref{c3}) one gets 
\begin{eqnarray}
a\,=\,\frac{-c^2\pm \sqrt{c^4-4b^2 c}}{2b}\end{eqnarray} 

The choices $b\,\,=\,\,c$, followed by $c\,=\,1$ give  the following ten matrices

\begin{eqnarray}\begin{array}{ccccc}
A_1=\left[\begin{array}{ccc}
\omega&1&1\\*[2mm]
1&\omega&1\\*[2mm]
1&1&\omega
\end{array}\right], &
A_2=\left[\begin{array}{ccc}
1&\omega&1\\*[2mm]
1&1&\omega\\*[2mm]
\omega&1&1
\end{array}\right],& 
A_3=\left[\begin{array}{ccc}
\omega &\omega&1\\*[2mm]
1&\omega &\omega\\*[2mm]
\omega &1&\omega
\end{array}\right],& 
A_4=\left[\begin{array}{ccc}
1&1&\omega\\*[2mm]
\omega&1&1\\*[2mm]
1&\omega&1 \end{array}\right],&
A_5=\left[\begin{array}{ccc}
\omega&1&\omega\\*[2mm]
\omega&\omega&1\\*[2mm]
1&\omega&\omega\end{array}\right]\label{a15}\end{array}\end{eqnarray}
and respectively 
\begin{eqnarray}\begin{array}{ccccc}
B_{1}= \left[\begin{array}{ccc}
 \omega^2&1&1\\*[2mm]
1&\omega^2&1\\*[2mm]
1&1&\omega^2
\end{array}\right],&
B_2=\left[\begin{array}{ccc}
1&\omega^2&1\\*[2mm]
1&1&\omega^2\\*[2mm]
\omega^2&1&1
\end{array}\right],& 
B_3=\left[\begin{array}{ccc}
\omega^2 &\omega^2&1\\*[2mm]
1&\omega^2 &\omega^2\\*[2mm]
\omega^2 &1&\omega^2
\end{array}\right],& 
B_4=\left[\begin{array}{ccc}
1&1&\omega^2\\*[2mm]
\omega^2&1&1\\*[2mm]
1&\omega^2&1 
\end{array}\right],&
B_5=\left[\begin{array}{ccc}
\omega^2&1&\omega^2\\*[2mm]
\omega^2&\omega^2&1\\*[2mm]
1&\omega^2&\omega^2\end{array}\label{b15}\right]\end{array}\end{eqnarray}

All the above ten matrices are not Hadamard
\section{New Hadamard Matrices}

The $A_1$ matrix generates two complex Hadamard matrices
 \begin{eqnarray}\begin{array}{cc}
A_{11}=\left[\begin{array}{ccc}
-(-1)^{1/3} &1&1\\*[2mm]
1&-(-1)^{1/3} &1\\*[2mm]
1&1&-(-1)^{1/3}\end{array}\right],&
A_{12}=\left[\begin{array}{ccc}
(-1)^{2/3}&1&1\\*[2mm]
1&(-1^{2/3}&1\\*[2mm]
1&1&(-1)^{2/3}
\end{array}\label{a1112}\right]\end{array}\end{eqnarray}
and $A_{11}$ and  $A_{12}$ lead to a MUB set  as ({$\mathbb{I}$, $A_{11}$, $A_{12}$}).
The matrices  $A_{11}$ and  $A_{12}$, and respectively  $A_{12}$ and  $A_{11}$ generate the following two  matrices
\begin{eqnarray}\begin{array}{cc}
A_{1112}=\left[\begin{array}{ccc}
(-1)^{1/6}&{-\bf i}&{-\bf i}\\*[2mm]
{-\bf i}&(-1)^{1/6}&{-\bf i}\\*[2mm]
{-\bf i}&{-\bf i}&(-1)^{1/6}
\end{array}\right],&
A_{1211}=\left[\begin{array}{ccc}
(-1)^{5/6}&-{\bf i}&-{\bf i}\\*[2mm]
-{\bf i}&(-1)^{5/6}&-{\bf i}\\*[2mm]
{-\bf i}&{-\bf i}&(-1)^{5/6}
\end{array}\label{a11121211}\right]\end{array}\end{eqnarray}
In this case the MUB is ({$\mathbb{I}$, $A_{1112}$, $A_{1211}$})

The $A_2$ matrix generates also two complex Hadamard matrices
\begin{eqnarray}\begin{array}{cc}
A_{21}=\left[\begin{array}{ccc}
1&-(-1)^{1/3} &1\\*[2mm]
1& 1&-(-1)^{1/3}\\*[2mm]
-(-1)^{1/3}&1&1\end{array}\right],&
A_{22}=\left[\begin{array}{ccc}
1&(-1)^{2/3}&1\\*[2mm]
1&1&(-1^{2/3}\\*[2mm]
(-1)^{2/3}&1&1
\end{array}\label{a2122}\right]\end{array}\end{eqnarray}
The matrices $A_{21}$ and  $A_{22}$ and respectively  $A_{22}$ and $A_{21}$ generate the same matrices (\ref{a11121211}).

The $A_3$ matrix generates two complex Hadamard matrices
\begin{eqnarray}\begin{array}{cc}
A_{31}=\left[\begin{array}{ccc}
-(-1)^{1/3}&-(-1)^{1/3}&1\\*[2mm]
1&-(-1)^{1/3}&-(-1)^{1/3} \\*[2mm]
-(-1)^{1/3}&1&-(-1)^{1/3} \end{array}\right],&
A_{32}=\left[\begin{array}{ccc}
(-1)^{2/3}&(-1)^{2/3}&1\\*[2mm]
1&(-1^{2/3}& (-1^{2/3}\\*[2mm]
(-1)^{2/3}&1& (-1^{2/3})
\end{array}\label{a3132}\right]\end{array}\end{eqnarray}
 Matrices $A_{31}$ and  $A_{32}$ lead to the MUB set ({$\mathbb{I}$, $A_{31}$, $A_{22}$}).
Similar to the preceding cases matrices  $A_{31}$ and  $A_{32}$ in this order, and respectively $A_{32}$ and  $A_{31}$ generate the matrices 
\begin{eqnarray}\begin{array}{cc}
A_{3132}=\left[\begin{array}{ccc}
{\bf i}&-(-1)^{1/6}&-(-1)^{1/6}\\*[2mm]
-(-1)^{1/6}&{\bf i}&-(-1)^{1/6}\\*[2mm]
-(-1)^{1/6}&-(-1)^{1/6}&{\bf i}
\end{array}\right],&
A_{3231}=\left[\begin{array}{ccc}
{\bf i}&-(-1)^{5/6}&-(-1)^{5/6}\\*[2mm]
-(-1)^{5/6}&{\bf i}&-(-1)^{5/6}\\*[2mm]
-(-1)^{5/6}&-(-1)^{5/6}&{\bf i}
\end{array}\label{A31323231}\right]\end{array}\end{eqnarray}
The above matrices generate the MUB ({$\mathbb{I}$, $A_{3132}$, $A_{3212}$}).

The matrices  $A_{1112}$ and  $A_{3122}$, and respectively $A_{3122}$ and  $A_{1112}$ generate the following matrices
\begin{eqnarray}\begin{array}{cc}
D_{11}=\left[\begin{array}{ccc}
(-1)^{1/6}&(-1)^{5/6}&(-1)^{5/6}\\*[2mm]
(-1)^{5/6}&(-1)^{1/6}&(-1)^{5/6}\\*[2mm]
(-1)^{5/6}&(-1)^{5/6}&(-1)^{1/6}
\end{array}\right],&
D_{12}=\left[\begin{array}{ccc}
(-1)^{5/6}&(-1)^{1/6}&(-1)^{1/6}\\*[2mm]
(-1)^{1/6}&(-1)^{5/6}&(-1)^{1/6}\\*[2mm]
(-1)^{1/6}&(-1)^{1/6}&(-1)^{5/6}
\end{array}\label{d11d12}\right]\end{array}\end{eqnarray}
Thus the MUB is given by ({$\mathbb{I}$, $D_{11}$, $D_{12}$}).

The $A_4$ matrix generates other two matrices whose form is
\begin{eqnarray}\begin{array}{cc}
A_{41}=\left[\begin{array}{ccc}
1&1&-(-1)^{1/3} \\*[2mm]
-(-1)^{1/3}&1&1\\*[2mm]
1&-(-1)^{1/3}&1
\end{array}\right],&
A_{42}=\left[\begin{array}{ccc}
1&1&(-1)^{2/3} \\*[2mm]
(-1)^{2/3}&1&1\\*[2mm]
1&(-1)^{2/3}&1
\end{array}\label{a4142}\right]\end{array}\end{eqnarray}
and the MU pair has the form ({$\mathbb{I}$, $A_{41}$, $A_{42}$}).
The matrices $A_{41}$ and  $A_{42}$, and respectively $A_{42}$ and $A_{41}$   generate again  the  matrices (\ref{a11121211}).

In the next case $A_5$ matrix leads also  to two matrices 
\begin{eqnarray}\begin{array}{cc}
A_{51}=\left[\begin{array}{ccc}
-(-1)^{1/3}&1&-(-1)^{1/3} \\*[2mm]
-(-1)^{1/3}&-(-1)^{1/3}&1\\*[2mm]
1&-(-1)^{1/3}&-(-1)^{1/3}
\end{array}\right],&
A_{52}=\left[\begin{array}{ccc}
(-1)^{2/3}  &1&(-1)^{2/3} \\*[2mm]
(-1)^{2/3}&(-1)^{2/3} &1\\*[2mm]
1&(-1)^{2/3}&(-1)^{2/3}
\end{array}\label{a5152}\right]\end{array}\end{eqnarray}
with the MUB pair written as  ($\mathbb{I}$, $A_{51}$, $A_{52}$). The matrices generated  $A_{51}$, and $A_{52}$ coincide with the matrices (\ref{a1112}).

The matrices $A_{51}$, $A_{52}$ generate the following matrices
\begin{eqnarray}\begin{array}{cc}
A_{5152}=\left[\begin{array}{ccc}
\bf i&-(-1)^{1/6}&-(-1)^{1/6}\\*[2mm]
-(-1)^{1/6}&\bf i&-(-1)^{1/6}\\*[2mm]
-(-1)^{1/6}&-(-1)^{1/6}&{\bf i}
\end{array}\right],&
A_{5251}=\left[\begin{array}{ccc}
\bf i&-(-1)^{5/6}&-(-1)^{5/6}\\*[2mm]
-(-1)^{5/6}&\bf i&-(-1)^{5/6}\\*[2mm]
-(-1)^{5/6}&-(-1)^{5/6}&{\bf i}
\end{array}\label{A5152a5251}\right]\end{array}\end{eqnarray}
and the MUB is  ({$\mathbb{I}$, $A_{5152}$, $A_{5251}$}).

With the $B_i$ matrices  one get similar results. Thus the  $B_1$ matrix leads to the following diagonal matrices
\begin{eqnarray}\begin{array}{cc}
B_{11}=\left[\begin{array}{ccc}
(-1)^{2/3} &1&1\\*[2mm]
1&(-1)^{2/3} &1\\*[2mm]
1&1&(-1)^{2/3}\end{array}\right],&
B_{12}=\left[\begin{array}{ccc}
-(-1)^{1/3}&1&1\\*[2mm]
1&-(-1)^{1/3} &1\\*[2mm]
1&1&-(-1)^{1/3}
\end{array}\label{b1112}\right]\end{array}\end{eqnarray}
The MUB set is  ($\mathbb{I}$, $B_{11}$, $B_{12}$). Similar to the preceding cases  $B_{11}$ and  $B_{12}$ generate the matrices (\ref{A5152a5251}).

The $B_2$ matrix leads to 
\begin{eqnarray}\begin{array}{cc}
B_{21}=\left[\begin{array}{ccc}
(-1)^{2/3}&1&1\\*[2mm]
1&(-1)^{2/3}&1 \\*[2mm]
1&1&(-1)^{2/3}\end{array}\right],&
B_{22}=\left[\begin{array}{ccc}
-(-1)^{1/3}&1&1\\*[2mm]
1&-(-1)^{1/3}&1\\*[2mm]
1&1&-(-1)^{1/3}\end{array}\label{b21b22}\right]\end{array}\end{eqnarray}
and the MUB set is  ({$\mathbb{I}$, $B_{21}$, $B_{22}$}).
Matrices $B_{2122}$ and  $B_{2221}$ coincide with the matrices $A_{1112}$ and $A_{211}$.

The $B_3$ matrix generates other two matrices
\begin{eqnarray}\begin{array}{cc}
B_{31}=\left[\begin{array}{ccc}
(-1)^{2/3}&(-1)^{2/3} &1\\*[2mm]
1&(-1)^{2/3}&(-1)^{2/3}  \\*[2mm]
(-1)^{2/3}& 1&(-1)^{2/3} \end{array}\right],&
B_{32}=\left[\begin{array}{ccc}
-(-1)^{1/3}&-(-1)^{1/3} &1\\*[2mm]
1&-(-1)^{1/3}&-(-1)^{1/3}\\*[2mm]
-(-1)^{1/3}&1&-(-1)^{1/3}
\end{array}\label{b3132}\right]\end{array}\end{eqnarray}
The  MU set is  ({$\mathbb{I}$, $B_{31}$, $B_{32}$}).
Matrices $B_{31}$ and  $B_{32}$ generate the  matrices $A_{5152}$ and  $A_{5251}$.

The $B_4$ matrix generate the following two Hadamard matrices
\begin{eqnarray}\begin{array}{cc}
B_{41}=\left[\begin{array}{ccc}
1 &1&(-1)^{2/3}\\*[2mm]
(-1)^{2/3}&1&1\\*[2mm]
1&(-1)^{2/3}&1 \end{array}\right],&
B_{42}=\left[\begin{array}{ccc}
1&1&-(-1)^{1/3}\\*[2mm]
-(-1)^{1/3}&1&1\\*[2mm]
1&-(-1)^{1/3}&1
\end{array}\label{b4142}\right]\end{array}\end{eqnarray}
and the MU form is ({$\mathbb{I}$, $B_{41}$, $B_{42}$}).
The matrices $B_{41}$ and  $B_{42}$ generate the matrices 
\begin{eqnarray}\begin{array}{cc}
B_{4142}=\left[\begin{array}{ccc}
(-1)^{5/6}&-\bf i& -{\bf i} \\*[2mm]
-\bf i&(-1)^{5/6}&-{\bf i}\\*[2mm]
- {\bf i}&-{\bf i}&(-1)^{5/6}
\end{array}\right],&
B_{4241}=\left[\begin{array}{ccc}
(-1)^{1/6}&-\bf i&-{\bf i}\\*[2mm]
-\bf i&(-1)^{1/6}&{-\bf i}\\*[2mm]
-\bf i&-\bf i&(-1)^{1/6}
\end{array}\label{B41424241}\right]\end{array}\end{eqnarray}
The MUB is given by ({$\mathbb{I}$, $B_{4142}$, $B_{4241}$}).

As usually the $B_5$ matrix  generates  other two  matrices
\begin{eqnarray}\begin{array}{cc}
B_{51}=\left[\begin{array}{ccc}
(-1)^{2/3}&1&(-1)^{2/3} \\*[2mm]
(-1)^{2/3}&(-1)^{2/3}&1\\*[2mm]
1&(-1)^{2/3}&(-1)^{2/3} \end{array}\right],&
B_{52}=\left[\begin{array}{ccc}
-(-1)^{1/3}&1&-(-1)^{1/3}\\*[2mm]
-(-1)^{1/3}&-(-1)^{1/3}&1\\*[2mm]
1&-(-1)^{1/3}&-(-1)^{1/3}
\end{array}\label{B5152}\right]\end{array}\end{eqnarray}
The MUB pair is  ({$\mathbb{I}$, $B_{51}$, $B_{52}$}).

The matrices $A_{1112}$ and  $B_{5152}$ generate the unitary diagonal matrix 
\begin{eqnarray}\begin{array}{c}
I_{2/3}=\left[\begin{array}{ccc}
(-1)^{2/3}&0&0\\*[2mm]
0&(-1)^{2/3}&0\\*[2mm]
0&0&(-1)^{2/3}
\end{array}\label{I_{2/3}}\right]\end{array}\end{eqnarray}
The correponding MUB have the form ($\mathbb{I}_{2/3}$,\,$A_{1112}$\, $B_{5152}$).
The same matrix is generated  by  $A_{4142}$ and  $B_{5152}$, etc.

The matrices $B_{3132}$ and  $A_{1112}$ generate another unitary diagonal matrix
\begin{eqnarray}\begin{array}{c}
I_{1/3}=\left[\begin{array}{ccc}
-(-1)^{1/3}&0&0\\*[2mm]
0&-(-1)^{1/3}&0\\*[2mm]
0&0&-(-1)^{1/3}
\end{array}\label{I_{m1/3}}\right]\end{array}\end{eqnarray}
 The MUB in this case has the form  ($\mathbb{I}_{m1/3}$,\,\,$B_{5152}$\,\,$A_{1112}$), where $m\,=\,-1$. The same matrix is generated by $B_{5152}$ and $A_{2122}$, and respectively by  $B_{4122}$ and $A_{3132}$, etc.

Our approach has shown that the matrices $A_i$ and $B_i$, $i\,=\,1,\,2,\,3,\,4\,, 5$ are not complex Hadamard matrices. Thus the first step was to make use of the Sylvester trick, see equation (\ref{di}), in order to transform all the matrices which depend on $\omega$ and $\omega^2$ into true Hadamard matrices. The final result is that with them we found many new MU bases.

\section{The five dimensional case}

This case is very interesting because there is a 5-dimensional circulant matrix of the following form 
 \begin{eqnarray}
C_5= \left[\begin{array}{ccccc}
1&a&a^4&a^4&a\\*[2mm]
a&1&a&a^4&a^4\\*[2mm]
a^4&a&1&a&a^4\\*[2mm]
a^4&a^4&a&1&a\\*[2mm]
a&a^4&a^4&a&1
\end{array}\right]\label{c5}\end{eqnarray}
which leads to new complex Hadamard matrices. For this we make use of the relation (\ref{di})for getting  complex Hadamard matrices from $C_5$ matrix. The diagonal entries are equal to 1. The off diagonal entries contain the polynomial $1+a+a^2+a^3+a^4$

For that we make use of the Sylvester relation \cite{JJS}, and 
the resulting matrix has 1 on the main diagonal and   all the other entries have a common factor given by
\begin{eqnarray}
1+a+a^2+a^3+a^4\label{fac}
\end{eqnarray}
The solutions of (\ref{fac}) give 5-dimensional matrices, and they are
\begin{eqnarray}
sol=\left[\begin{array}{cccc}
a_1 = -1(-1)^{1/5} , a_2 = (-1)^{2/5} , a_3 = -1(-1)^{3/5}, a_4 = (-1)^{4/5}\end{array}\right]\label{sol} \end{eqnarray}

By using each solution in matrix (\ref{c5}) we get four different matrices denoted by $D_i$, i= 1,2,3,4. They are

\tiny{
\begin{eqnarray}
D_1 = \left[\begin{array}{ccccc}
 1& -(-1)^{1/5}& (-1)^{4/5}&(-1)^{4/5}&-(-1)^{1/5}\\*[2mm]
 -(-1)^{1/5}&1&-(-1)^{1/5}&(-1)^{4/5}&(-1)^{4/5}\\*[2mm]
(-1)^{4/5}& -(-1)^{1/5}&1& -(-1)^{1/5}&(-1)^{4/5}\\*[2mm]
(-1)^{4/5}&(-1)^{4/5}& -(-1)^{1/5}&1&-(-1)^{1/5}\\*[2mm]
-(-1)^{1/5}&(-1)^{4/5}&(-1)^{4/5}&-(-1)^{1/5}&1
\end{array}\right],&
D_2 = \left[\begin{array}{ccccc}
1&(-1)^{2/5}& -(-1)^{3/5}& -(-1)^{3/5}&(-1)^{2/5}\\*[2mm]
(-1)^{2/5}&1&(-1)^{2/5}& -(-1)^{3/5}& -(-1)^{3/5}\\*[2mm]
-(-1)^{3/5}&(-1)^{2/5}&1&-(-1)^{2/5}&-(-1)^{3/5}\\*[2mm]
-(-1)^{3/5}&-(-1)^{3/5}&(-1)^{2/5}&1&(-1)^{2/5}\\*[2mm]
(-1)^{2/5}&-(-1)^{3/5}&-(-1)^{3/5}&(-1)^{2/5}&1
\end{array}\right]\label{d12}\end{eqnarray}}

\tiny{
\begin{eqnarray}
D_3 = \left[\begin{array}{ccccc}
1&-(-1)^{3/5}&(-1)^{2/5}&(-1)^{2/5}&-(-1)^{3/5}\\*[2mm]
-(-1)^{3/5}&1&-(-1)^{3/5}&(-1)^{2/5}&(-1)^{2/5}\\*[2mm]
(-1)^{2/5}&-(-1)^{3/5}& 1&-(-1)^{3/5}&(-1)^{2/5}\\*[2mm]
(-1)^{2/5}&(-1)^{2/5}&-(-1)^{3/5}&1&-(-1)^{3/5}\\*[2mm]
-(-1)^{3/5}&(-1)^{2/5}&(-1)^{2/5}&-(-1)^{3/5}&1
\end{array}\right],&
D_4 =\left[\begin{array}{ccccc}
1&(-1)^{4/5}&-(-1)^{1/5}&-(-1)^{1/5}&(-1)^{4/5}\\*[2mm]
(-1)^{4/5}&1&(-1)^{4/5}&-(-1)^{1/5}&-(-1)^{1/5}\\*[2mm]
-(-1)^{1/5}&(-1)^{4/5}&1&(-1)^{4/5}&-(-1)^{1/5}\\*[2mm]
-(-1)^{1/5}&-(-1)^{1/5}&(-1)^{4/5}&1&(-1)^{4/5}\\*[2mm]
(-1)^{4/5}&-(-1)^{1/5}&-(-1)^{1/5}&(-1)^{4/5}&1
\end{array}\right]\label{d34}\end{eqnarray}}
\normalsize
All the above four matrices are complex Hadamard and they generate a MUB of the following  form ($\mathbb{I},D_1,D_2,D_3,D_4$).

\section{Conclusion}

Our approach has shown that the matrices $A_i$ and $B_i$ are not complex Hadamard. So we used the necessary constraints  to obtain  new Hadamard matrices. In the same time we disproved 
the assertion that for  3- and 5-dimensional matrices they have the form of the corresponding Fourier matrices. An important result is that $C_5$ matrix generated four Hadamard matrices which disprove all assumptions  rised  in papers  \cite{BF},  \cite{D} and \cite{S}.

\end{document}